\documentclass[apj]{emulateapj}
\usepackage{amsmath}%
\usepackage{amssymb}%
\usepackage{srcltx}
\usepackage[english]{babel}
\usepackage{hyperref}
\usepackage{graphicx}

\begin{document}
\selectlanguage{english}

\title{The Outburst of the Blazar \object{S5~0716+71} in 2011 October: Shock in a Helical Jet}
\author{V.M.\,Larionov\altaffilmark{1,3}, 
S.G.\,Jorstad\altaffilmark{1,2}, A.P.\,Marscher\altaffilmark{2}, 
D.A.\,Morozova\altaffilmark{1}, D.A.\,Blinov\altaffilmark{1}, V.A.\,Hagen-Thorn\altaffilmark{1,3}, T.S.\,Konstantinova\altaffilmark{1}, E.N.\,Kopatskaya\altaffilmark{1},  L.V.\,Larionova\altaffilmark{1}, E.G.\,Larionova\altaffilmark{1},  I.S.\,Troitsky\altaffilmark{1}
}

\altaffiltext{1}{Astronomical Institute of St. Petersburg State University, Universitetsky Pr. 28, Petrodvorets, 198504, St. Petersburg, Russia; vlar@astro.spbu.ru}
\altaffiltext{2}{Institute for Astrophysical Research, Boston University,
725 Commonwealth Ave., Boston, MA 02215-1401; jorstad@bu.edu}
\altaffiltext{3}{Isaac Newton Institute of Chile, St.-Petersburg Branch}
\shorttitle{Outburst of S5~0716+71 in 2011 October.}
\shortauthors{Larionov et al.}
\begin{abstract}
We present the results of optical ($R$ band) photometric and polarimetric monitoring and Very Long Baseline Array (VLBA) imaging of the blazar \object{S5~0716+714} along with \textit{Fermi} $\gamma$-ray data during a multi-waveband outburst in 2011 October. We analyze total and polarized intensity images of the blazar obtained with the VLBA at 43 GHz during and after the outburst. Monotonic rotation of the linear polarization vector at a rate of $\ga 50\degr$ per night coincided with a sharp maximum in $\gamma$-ray and optical flux. At the same time, within the uncertainties, a new superluminal knot
appeared, with an apparent speed of $21\pm 2~c$. The general multi-frequency behavior of the outburst can be explained within the framework of a shock wave propagating along a helical path in the blazar's jet. 
\end{abstract}

\section{Introduction}
The blazar \object{S5~0716+71} is one of most intensively studied BL Lac objects. The redshift determinations ($z\approx 0.3$) are based on measurements of the size of a marginally detected  host galaxy and a location near three galaxies with redshifts of 0.26 \citep{2006ARep...50..802B, 2008A&A...487L..29N}. Optical variability has been studied by many teams; the blazar exhibits persistent activity on both long and short (intraday) timescales, as shown by, e.g., \citet{1996AJ....111.2187W}; \citet{1997A&A...327...61G}, \citet{2006ARep...50..458H}, and \citet{2006MNRAS.366.1337S,2009MNRAS.399.1357S}, and by the results of several WEBT campaigns \citep{2003A&A...402..151R,2006A&A...451..797O,2008A&A...481L..79V,2008A&A...489L..37C}. Violent polarimetric variability of S5~0716+71, among other blazars, was studied by \citet{2011PASJ...63..639I}.  \citet{2010PASJ...62...69U}, using a Bayesian approach, suggest that the behavior of polarization in S5~0716+714  implies the presence of a number of polarization components having quite short timescales of variations. \citet{2008ATel.1502....1L} reported that a huge optical outburst of S~50716+71 in 2008 April was accompanied by a $\sim 360\degr$ rotation of the position angle of the electric vector (EVPA) of linear polarization.

In this paper, we report the results of our observations of S5~0716+71 during a major optical outburst in 2011 October, which coincided with an unprecedented $\gamma$-ray outburst \citep{2011ATel.3700....1B}. We present our observations in \S~\ref{obs}, describe the phenomenological model in \S~\ref{sec:model}, and apply it to the data and draw conclusions in \S~\ref{discuss}.

\section{Observations and Results}\label{obs}
The observations reported here were collected as a part of a long-term multi-wavelength study of blazars that follows nonthermal activity at optical and $\gamma$-ray frequencies. An overview of this program is given in \citet{2012IJMPS...8..151M}.

\subsection{Optical Observations}
We obtained observational data at 
the 70-cm AZT-8 reflector of the Crimean Astrophysical Observatory, 40-cm LX-200 telescope in St.~Petersburg, both equipped with identical photometers-polarimeters based on ST-7 CCDs, and 1.8-m Perkins telescope of Lowell Observatory using the PRISM camera. The photometric measurements in $R$ band were supplemented by observations at the 2-m Liverpool Telescope at La Palma, Canary Islands, Spain. Polarimetry at AZT-8 and the Perkins telescope was made in Cousins $R$ band, and in white light at LX-200, with effective wavelength close to $R$. The photometric errors do not exceed $0\fm02$.

Instrumental polarization was found via stars located near the object under the assumption that their radiation is unpolarized. The Galactic latitude of S5~0716+714 is $28\degr$ and $A_V=0\fm1$, so that interstellar polarization (ISP) in this direction is less than 0.5\%. To correct for the ISP, the mean relative Stokes parameters of nearby stars were subtracted from the relative Stokes parameters of the object. This accounts for the instrumental polarization as well. The errors in degree of polarization are less than 1\% (in most cases less than 0.5\%), while the EVPA is determined with a precision of 1--2 degrees.

\subsubsection{Results}
Our regular photometric and polarimetric monitoring of S5~0716+71 started in 2005. Since then, we have obtained more than 1000 data points almost uniformly spread over a 7-year interval, although we intensified our observations during the periods of highest activity of the source.

\begin{figure}[ht]
\begin{center}
   \includegraphics[width=7.5cm,clip]{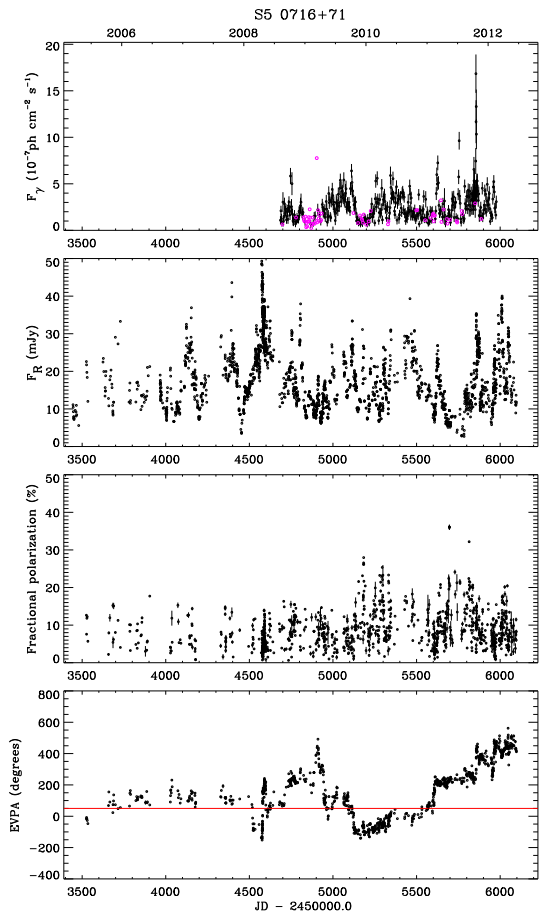}
        \caption{From top to bottom: evolution of $\gamma$-ray flux density (purple points mark upper limits), $R$-band flux, and degree and position angle of optical polarization of S5~0716+71 during 2005--2011. Red horizontal line in the bottom panel corresponds to the position angle of the radio jet in 2011.}
        
\label{s5total}
\end{center} 
\end{figure}

\begin{figure}[ht]
\begin{center}
   \includegraphics[width=7.5cm,clip]{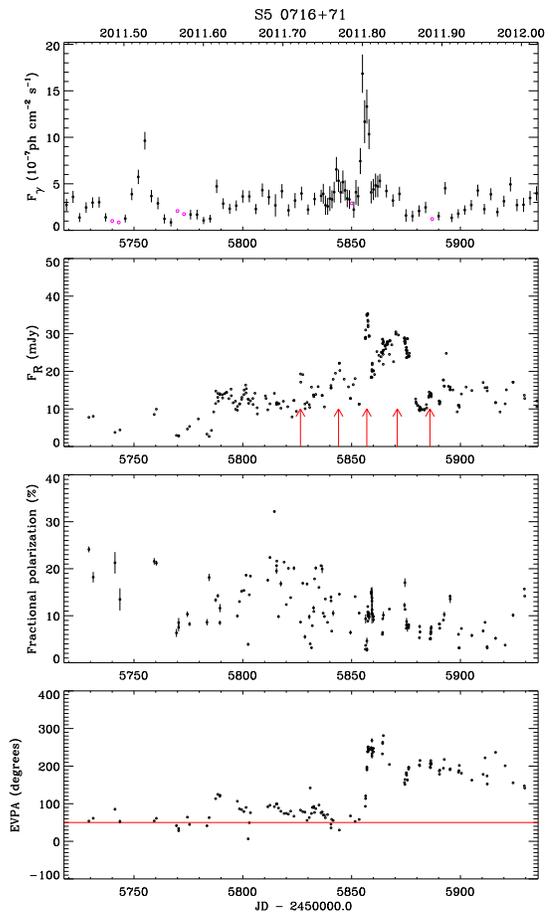}
        \caption{Blowup of Fig~\ref{s5total} centered around 2011 outburst. Red arrows on $R$ flux panel mark positions of quasi-periodic optical flares.}
        
\label{s5total2011}
\end{center} 
\end{figure}

Figure~\ref{s5total} presents the flux and polarization behavior of \object{S5~0716+71} for 2005--2011. We supplement this plot with a panel showing the $\gamma$-ray light curve from \textit{Fermi} LAT open access data in order to show that the most prominent $\gamma$-ray activity ever recorded for this source was observed during the October 2011 optical outburst. In Fig.~\ref{s5total2011} we show a blowup of the last six months of 2011. 
Table~1 lists the photometric and polarimetric data of S5~0716+71 around the time of the outburst.

\begin{deluxetable*}{cccrcccc}

%% Keep a portrait orientation

%% Over-ride the default font size
%% Use Default (12pt)

%% Use \tablewidth{?pt} to over-ride the default table width.
%% If you are unhappy with the default look at the end of the
%% *.log file to see what the default was set at before adjusting
%% this value.

%% This is the title of the table.
\tablecaption{Photometry and polarimetry of S5~0716+71 during 2011 October outburst}

%% This command over-rides LaTeX's natural table count
%% and replaces it with this number.  LaTeX will increment 
%% all other tables after this table based on this number
%\tablenum{1}

%% The \tablehead gives provides the column headers.  It
%% is currently set up so that the column labels are on the
%% top line and the units surrounded by ()s are in the 
%% bottom line.  You may add more header information by writing
%% another line between these lines. For each column that requries
%% extra information be sure to include a \colhead{text} command
%% and remember to end any extra lines with \\ and include the 
%% correct number of &s.
\tablehead{\colhead{JD-2400000} & \colhead{$R$} & \colhead{$\sigma R$} & \colhead{$p$} & \colhead{$\sigma p$} & \colhead{$EVPA$} & \colhead{$\sigma EVPA$} & \colhead{Telescope} \\ 
\colhead{(days)} & \colhead{(mag)} & \colhead{(mag)} & \colhead{(\%)} & \colhead{(\%)} & \colhead{($\degr$)} & \colhead{($\degr$)} & \colhead{} } 

%% All data must appear between the \startdata and \enddata commands
\startdata
55823.516 & 13.496 & 0.012 & 20.11 & 0.41 & 246.7 & 0.5 & LX-200   \\
55826.543 & 13.002 & 0.003 & 8.64 & 0.17 & 262.9 & 0.5 & AZT-8+ST7 \\
55827.554 & 13.012 & 0.002 & 16.90 & 0.24 & 258.7 & 0.4 & AZT-8+ST7 \\
55828.547 & 13.582 & 0.003 & 5.51 & 0.40 & 257.6 & 2.1 & AZT-8+ST7 \\
55829.544 & 13.542 & 0.004 & 16.75 & 0.28 & 235.6 & 0.4 & AZT-8+ST7 \\
55830.560 & 13.598 & 0.003 & 9.77 & 0.50 & 243.0 & 1.4 & AZT-8+ST7 \\
55831.574 & 13.219 & 0.002 & 3.17 & 0.16 & 254.0 & 1.5 & AZT-8+ST7 \\
55832.531 & 13.373 & 0.003 & 11.66 & 0.41 & 269.8 & 1.0 & LX-200   \\
55833.533 & 13.361 & 0.003 & 20.14 & 0.32 & 266.9 & 0.4 & AZT-8+ST7 \\
55836.488 & 13.384 & 0.003 & 19.93 & 0.87 & 258.1 & 1.2 & LX-200   \\
\enddata

%% Include any \tablenotetext{key}{text}, \tablerefs{ref list},
%% or \tablecomments{text} between the \enddata and 
%% \end{deluxetable} commands

%% General table comment marker
\tablecomments{Table 1 is published in its entirety in the electronic edition of the Journal. A portion is shown here for guidance regarding its form and content.}

%% General table references marker
%\tablerefs{observations}

\end{deluxetable*}

The optical light curve of S5~0716+71 displays violent variability on both long (months-years) and short (days-weeks) timescales. Changes in the degree of polarization seem to occur erratically: within the time interval 2005--2009 it varied from 0 to 15\%, while later the mean level of polarization increased, reaching a record value of 36\% on the night of 2011 May 14 (JD~2455696). The EVPA also exhibited prominent changes. To solve the $\pm180\degr$ ambiguity, we have added/subtracted $180\degr$ each time that the subsequent value of EVPA is $> 90\degr$ less/more then the preceding one.

\subsection{Radio Observations}
The BL Lac object S5~0716+71 is monitored by the Boston University group with the Very Long Baseline Array (VLBA) at 43~GHz within a sample of bright $\gamma$-ray blazars\footnote{\url{http://www.bu.edu/blazars}} with an angular resolution of $\sim$0.1~mas,
which corresponds to 0.45~pc at the source distance ($H_0=70$, $\lambda_m=0.3$, $\lambda_\Omega=0.7$). The VLBA data were calibrated and imaged in the same manner as discussed in \citet{2005AJ....130.1418J}. 

\begin{figure}[ht]
\begin{center}
   \includegraphics[width=7.5cm,clip]{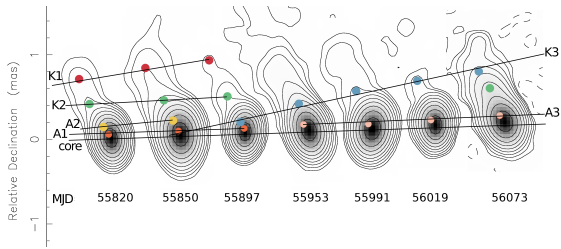}
        \caption{Uniformly weighted VLBA images at 43 GHz of moving emission knots in the S5~0716+071 jet. Total intensity contours are in factors of 2 starting at 0.25\%. The restoring beam is
identical to that shown in Figure \ref{0716_pfmaps}. New superluminal knot K3 appeared on
MJD~55850, which coincides within 1-$\sigma$ uncertainty with the time of the highest peak in the
$\gamma$-ray light curve.}
\label{comp_int}
\end{center} 
\end{figure}

\begin{figure*}[ht]
\begin{center}
   \includegraphics[width=15cm,clip]{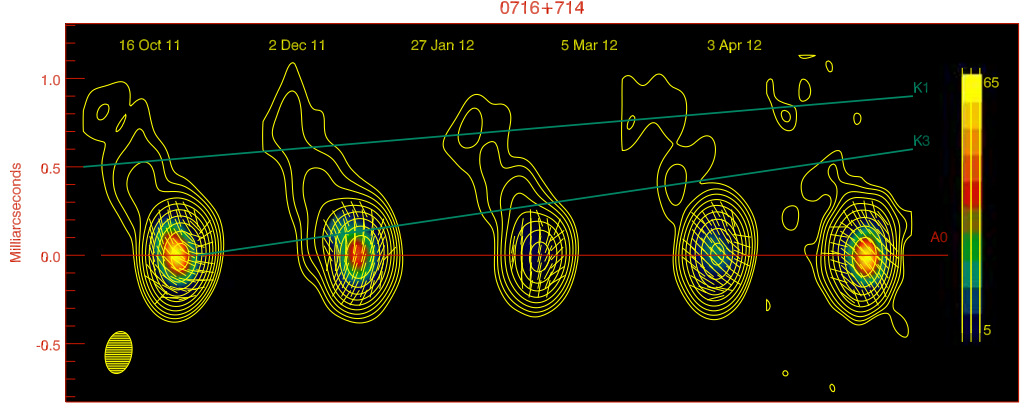}

        \caption{Total (contours) and polarized (color scale) intensity maps of
S5~0716+714 at 43~GHz with a total intensity peak of 2.51~Jy/beam and polarized
intensity peak of 65~mJy/beam; contours represent 0.15, 0.30,...,76.8\% of the peak of
intensity; the color bar (right) gives the color code for the polarized intensity  from
5~mJy/beam to 65~mJy/beam. The elliptical gaussian beam is 0.24$\times$0.15~mas$^2$
(FWHM) at PA=-10$^\circ$ (plotted in the bottom left corner); yellow line segments indicate direction of polarization; red and blue solid lines correspond to approximate
positions of the core $A0$ and superluminal knots $K1$ and $K3$.}
\label{0716_pfmaps}
\end{center} 
\end{figure*}

\begin{figure}[ht]
\begin{center}
   \includegraphics[width=7.5cm,clip]{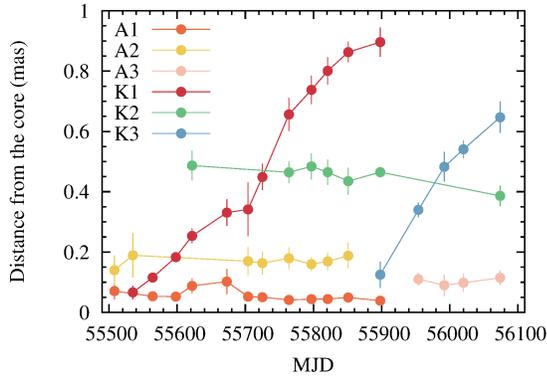}
        \caption{Position of jet features with respect to the core as a function of time.
}
\label{0716_components}
\end{center} 
\end{figure}

\begin{figure}[ht]
\begin{center}
   \includegraphics[width=7.5cm,clip]{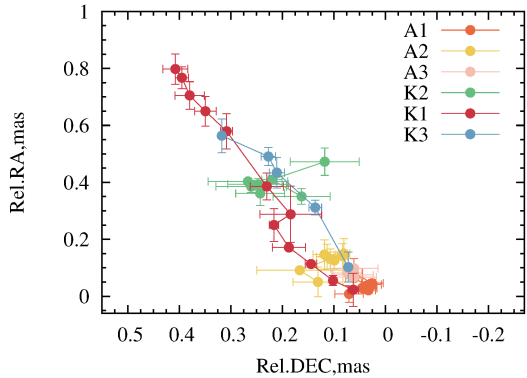}
        \caption{Trajectories of the jet knots.
}
\label{0716_traj}
\end{center} 
\end{figure}

We fit the VLBA data at 16 
epochs from 2010 December to 2012 May by a model consisting of a number of components with
circular Gaussian brightness distributions.
During this period we identify 6 components, $A1, A2, A3, K1, K2$, and $K3$, in addition 
to the core, $A0$ (Fig.~\ref{comp_int}), \ref{0716_pfmaps}), which is, presumably, a stationary feature located at the 
southern edge of the jet in the images. Figure~\ref{0716_components} shows the evolution of the 
distance of knots from the core, while Figure~\ref{0716_traj} plots the trajectories of the knots.

Knots $K1$ and $K3$ move to the northeast along slightly different trajectories
with apparent speeds of 18.1$\pm$0.5~c and 21$\pm$2~c, respectively. An apparent 
speed of $\sim$20~c seems to be common in the parsec scale jet of 
S5~0716+71 \citep[see, e.g.,][]{2011A&A...529A...2R}. Knot $K1$ undergoes both an acceleration 
along the jet $\mu_\|$=0.23$\pm$0.06~mas~yr$^{-2}$ and perpendicular to the jet 
$\mu_\bot$=-0.19$\pm$0.06~mas~yr$^{-2}$. 
An extrapolation of the motion of $K3$ back to the core suggests that it passed 
through the core on MJD~55850$\pm$10. Figure~\ref{0716_pfmaps} shows total and polarized
intensity images of the quasar while $K3$ emerged from the core.


Features $A1$, $A2$, $A3$, and $K2$ are rather stable in distance with respect to the core,
although they show scatter in position angle that exceeds the uncertainties of the modeling. 
\citet{2001ApJS..134..181J} found that ``stationary hot spots'' are a common characteristic
of compact jets,
with the majority of such features located within a range of projected distances of 1-3~pc 
from the core. These authors propose three categories of models for stationary components in 
supersonic jets: a) standing recollimation shocks caused by imbalances between the pressure
internal and external to the jet; b) sites of maximum Doppler beaming where the jet points most 
closely to the line of sight; and c) stationary oblique shocks, where the jet bends abruptly.
We conclude that knots $A1$, $A2$, and $A3$ fall most likely in category $a$, since 
they are quasi-stationary with an observed ``lifetime'' of several months to a year.
In addition, according to numerical simulations by \cite{1997ApJ...482L..33G}, when a
moving knot passes 
through a standing recollimation shock, the components blend into a single feature and split up
after the collision with no changes in proper motion of the moving knot. This scenario 
seems to be observed in the case of $K1$ approaching $A2$ (Fig.~\ref{0716_components}).
Knot $A3$, however, could be a trailing component that forms behind a strong shock
\citep{2001ApJ...549L.183A}, since it is observed just after the appearance of the fast moving
knot $K3$. The latter might change the internal pressure in the jet, resulting in the disappearance
of knots $A1$ and $A2$
(see Fig.~\ref{0716_components}). Knot $K2$ may belong to category $c$, since the position angle 
of $K2$ is different  
from that of $A1$, $A2$, and $A3$ (Fig.~\ref{0716_traj}) and the kinematics of moving knot $K1$
changes in the vicinity of $K2$. In the case of a stationary 
feature produced at a bend in the jet, the proper motion of a moving knot approaching the stationary 
feature should decrease before and increase after the knot passes through the bend \citep{1993ApJ...402..160A}. 
Such a behavior can be inferred from Figure~\ref{0716_components} when knot $K1$ approaches and 
recedes from knot $K2$.

\subsection{Gamma-ray Observations}
The $\gamma$-ray emission from S5~0716+71 was detected throughout the entire period of \textit{Fermi} observations starting in 2008 August, although most of the time at a level $\le 5\times10^{-7}\mbox{ph}\cdot \mbox{cm}^{-2}\mbox{s}^{-1}$. We analyze the \textit{Fermi} LAT data for the period of 2011 January-November with 2-day binning, as well as with 1-day binning and a 6-hour shift, during the huge $\gamma$-ray outburst of 2011 October, when the flux from the source was close to $2\times10^{-6}~\mbox{ph}\cdot \mbox{cm}^{-2}\mbox{s}^{-1}$. This prevents missing any possible short-lived events, to make the correlation
analysis more robust, and to avoid the dependence of the results on the start of
the time bins. We use the standard \textit{Fermi} analysis software package Science Tools
\texttt{v9r27p1}, with the instrument response function \texttt{P7SOURCE\_V6}, the Galactic
diffuse emission model \texttt{gal\_2yearp7v6\_v0}, and the isotropic background model
\texttt{iso\_p7v6source}. The background models include all sources from the 2FGL within
$15^\circ$. The spectrum of S5~0716+71 as well as spectra of all background
sources are modelled as a power-law with photon index fixed to the catalog
values.

\begin{figure}[ht]
\begin{center}
   \includegraphics[width=7.5cm,clip]{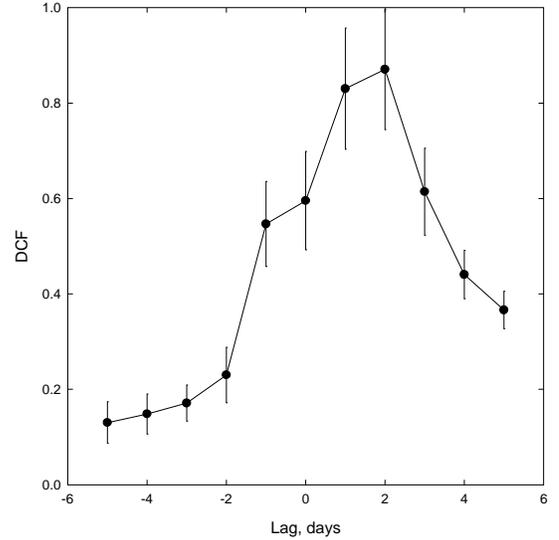}
        \caption{Discrete correlation function between optical and $\gamma$-ray flux variations of S5~0716+714. Positive lag means that optical variations follow those at $\gamma$-ray energies.}
\label{dcf}
\end{center} 
\end{figure}

We calculate the discrete correlation function (DCF) \citep{1988ApJ...333..646E} of the optical and $\gamma$-ray flux variations during 2011. The results are given in Fig.~\ref{dcf}. Optical variations lag those at $\gamma$-ray energies by $\approx1.4$ days (DCF centroid position). 

\section{Modeling of Photometric and Polarimetric Behavior}\label{sec:model}

We suppose, following \citet{2008Natur.452..966M}, that most of optical photometric and polarimetric variability arises when a compact emission region (e.g., a shock wave) propagates downstream from the black hole, following a spiral path. (Alternatively, the jet could have a helical geometry.) 
We notice that  prominent optical outbursts are quite often both preceded and followed by minor flares. We interpret this phenomenon as a manifestation of the oscillating Doppler beaming of the emission (lighthouse effect, see \citet{1992A&A...255...59C}). The observed series of outbursts correspond to the time intervals when the viewing angle of the shock wave is at a minimum.

The main parameters that determine the visible  behavior of the outburst are:

\begin{itemize}
\item{jet viewing angle $\theta$}
\item{bulk Lorentz factor $\Gamma$ of the shocked plasma}
\item{temporal evolution of the outburst $F'=F_0\cdot\left\{^{\exp(-|(t-t_0)|/\tau),\mbox{   }t\leq t_0}_{\exp(-|(t-t_0)|/k\tau),\mbox{  }t> t_0}\right.$; the factor $k$ is responsible for different time scales of the rise and decline of the outburst; primed quantities refer to the plasma frame}
\item{Doppler time contraction in the observer's frame, $\Delta t_{\mathrm{obs}}=\delta^{-1}\cdot \Delta t_\mathrm{src}$}
\item{shocked plasma compression $\eta$ = ratio of post-shock to pre-shock density}
\item{spectral index of the emitting plasma $\alpha$}
\item{pitch angle $\zeta$ of the spiral motion and helical field}
\item{period of the shock's spiral revolution in the observer's frame $P_\mathrm{obs}$}.
\end{itemize}

The viewing angle of the compact emission region $\varphi$ is obtained from the relation:

\begin{equation}
\cos\varphi=\cos\theta\cos\zeta+\sin\theta\sin\zeta\cos (2\pi t_{\mathrm{obs}}/P_{\mathrm{obs}}).
\label{eq1}
\end{equation}
Relativistic aberration leads to a change in direction of the normal to the shock front $\psi$:

\begin{equation}
\psi=\arctan[\sin\varphi/(\Gamma(\cos\varphi-\sqrt{1-\Gamma^{-2}}))]. \label{eq2}
\end{equation}

If the \textit{unshocked} magnetic field is chaotic upstream of the shock wave, the fractional polarization of the shocked plasma radiation \citep[see, e.g.,][]{Hughes1985} is:

\begin{equation}
p\approx \frac{\alpha+1}{\alpha+5/3}~\frac{(1-\eta^{-2})\sin^2\psi}{2-(1-\eta^{-2})\sin^2\psi}. \label{eq3}
\end{equation}

\noindent and the position angle is determined by the direction of the projected minor axis of the shock wave:

\begin{equation}
\Theta= \arctan[\zeta\sin t_\mathrm{obs}/(\zeta\cos t_\mathrm{obs}-\theta)]. 
\label{eq4}
\end{equation}

The Doppler factor $\delta$ is given by
\begin{equation}
\delta= 1/(\Gamma(1-\beta\cos\varphi)). 
\label{eq5}
\end{equation}

The periods of rotation in the source frame $P_\mathrm{src}$ and observer's frame $P_\mathrm{obs}$ are connected through the relation
\begin{equation}
P_\mathrm{obs}=P_\mathrm{src}(1+z)(1-\beta\cos\theta\cos\zeta),
\label{eq6}
\end{equation}
\noindent where $z$ is the source's redshift.

Knowledge of $P_\mathrm{obs}$ (from the timescale of recurrence of outbursts), estimates of $\beta$ and $\theta$, obtained from analysis of radio images, and evaluation of the spiral's radius $r$ allow us to constrain the values of $P_\mathrm{src}$ and $\zeta$ and to calculate the Doppler-contracted time in the observer's frame, $t_\mathrm{obs}$.
Doppler boosting of the radiation from the shocked plasma, as a discrete source, leads to
\begin{equation}
F_{\mathrm {shock}}= F'_{\mathrm {shock}}\delta^{3+\alpha}.
\label{eq10}
\end{equation}

We notice that in most cases when the source is in a quiescent state and no outburst is detected, there remains polarization at the level of $\leq 10\%$. We assume that this is the polarized radiation of the ambient jet, presumably changing on substantially longer timescales (except possibly for random fluctuations due to turbulence in the jet).
In the local frame the ratio $R1=F'_{\mathrm {shock}}/F'_{\mathrm{jet}}$ is of the order of a few percent. The radiation from the undisturbed jet is also subject to Doppler boosting, and we (rather arbitrarily) set $\Gamma_\mathrm{jet}\sim 0.7\Gamma$. Since the ambient jet is a roughly steady and continuous source, its radiation is amplified by $\delta^{2+\alpha}$. Notice that, in spite of the low value of $R1$, in the observer's frame the boosted radiation of the shock may far exceed that of the ambient jet.

We add emission from the quasi-stationary features --- probably standing shocks --- whose fluxes become significant after the moving shock crosses them. We postulate that the temporal behavior of this added radiation is of the same form as that of the shock: 
$$\Delta F'_{\mathrm{jet}}=R2\cdot F_0\cdot\left\{^{\exp(-|(t-t_0+t_\mathrm{del})|/\tau),\mbox{   }t\leq t_0}_{\exp(-|(t-t_0+t_\mathrm{del})|/k\tau),\mbox{  }t> t_0}\right.$$

Here $R2$ accounts for the ratio of the ambient jet to shock luminosity, with time delay $t_\mathrm{del}$. In our model we suppose that this radiation is added to the ``constant'' radiation of the jet, but not to its polarized part. Finally, the observed radiation of the jet is $F_\mathrm{jet}=(F'_\mathrm{jet}+\Delta F'_\mathrm{jet})\delta_\mathrm{jet}^{2+\alpha}$ and total observed flux  $F_\mathrm{total}=F_\mathrm{shock}+F_\mathrm{jet}$.

We consider a single variable source plus an initially constant source whose polarization vector is displaced relative to the jet direction by an angle $\Delta \chi$ \citep[see, e.g.,][]{2010LNP...794.....B}. 
Accounting for the contribution of this ``constant'' polarized source ($F'_\mathrm{jet}$ and $p_\mathrm{jet}$), we get the absolute Stokes parameters:

\begin{equation}
\begin{array}{l}
Q=F_\mathrm {shock}p\cos(2\Theta)+F'_\mathrm{jet}\delta_\mathrm{jet}^{2+\alpha}p_\mathrm{jet}\cos(2\Delta \chi)\\
U=F_\mathrm {shock}p\sin(2\Theta)+F'_\mathrm{jet}\delta_\mathrm{jet}^{2+\alpha}p_\mathrm{jet}\sin(2\Delta \chi).
\end{array}
\label{eq11}
\end{equation}

\noindent Finally, the normalized Stokes parameters are given by

\begin{equation}
\begin{array}{l}
q=Q/F_\mathrm{total}\\
u=U/F_\mathrm{total}.
\end{array}
\label{eq12}
\end{equation}

\begin{figure}[ht]
\begin{center}
   \includegraphics[width=7.5cm,clip]{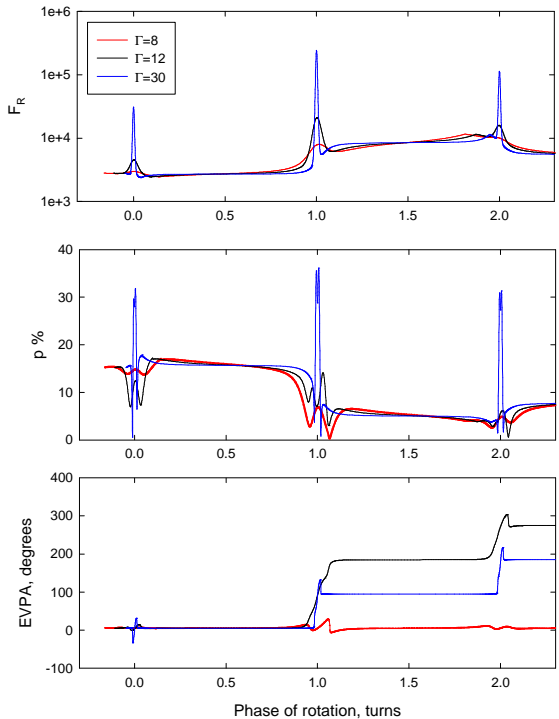}
        \caption{Model evolution of flux and polarization parameters with 3 different values of $\Gamma$.}
\label{model1}
\end{center} 
\end{figure}

Figure~\ref{model1} shows the dramatically different behavior of flux and polarization with all of the above mentioned parameters fixed, allowing only the Lorentz factor of the shock to change.
We see that the model of a shock propagating in the helical path may in a natural way explain the successions of outbursts, often observed in blazar light curves, as a manifestation of a single event. A variety of observed patterns of photometric and polarimetric behavior may be explained by the interplay of the shock Lorentz factor, jet viewing angle, helical path pitch angle, and shape and length of the outburst.  We do not take into account the light travel delay caused by the finite size of the shock wave; if included, this would result in some smoothing of the model light curve and
polarization behavior.

\section{Discussion and Conclusions}\label{discuss}
Quasi-periodic flares in the light curves of blazars are expected in models of magnetically launched jets, where the plasma moves with relativistic velocities following helical paths
\citep{2006ASPC..350..169V}. In light
of this, it is rather surprising that numerous extensive observational campaigns have uncovered only a few cases that could be interpreted as a manifestation of a lighthouse effect from spiral relativistic motion, e.g., \object{3C~273} \citep[$P=13^\mathrm d$, optical $V$ band][]{1988Natur.335..330C}; \object{OJ~287} \citep[$P=40^\mathrm d$, optical $VRI$ bands][]{2006AJ....132.1256W}; \object{BL~Lac} \citep[$P=308^\mathrm d$, optical photometry and polarimetry][]{2002AJ....124.3031H}; and {AO~0235+16} \citep[$P=17^\mathrm d$ in X-rays][]{2009ApJ...696.2170R}. Radio variability tends to occur on timescales longer than these periods. However, radio mapping provides some picturesque examples of periodic structures that correspond to longer timescales in AGN jets, as in the case of \object{PKS~0637-752} \citep[$2\times10^3$ years][]{2012ApJ...758L..27G}. As noted in \cite{2010A&A...510A..93L}, the jet might exhibit a fractal helical structure, where geometric scales that increase with distance from the central engine correspond to increasing variability timescales. Recent polarimetric studies of the nucleus of \object{M87} \citep{2011ApJ...743..119P} have revealed strong variations of the polarization vector, synchronous with the flux variations, that can be interpreted as a manifestation of a shock propagating through a helical jet.

We have obtained densely sampled optical photometric and polarimetric data during an unprecedented $\gamma$-ray outburst in S5~0716+71, and analyzed the changes in the structure of the blazar at 43~GHz.
We find a $180\degr$ rotation of the position angle of linear polarization, which occurred at the epoch of maximum optical brightness. This event would have been missed if our observations did not occur consecutively from 3 sites (St. Petersburg, Crimea, Arizona). This implies that such events may be rather common during blazar flares.

\begin{figure}[ht]
\begin{center}
   \includegraphics[width=7.5cm,clip]{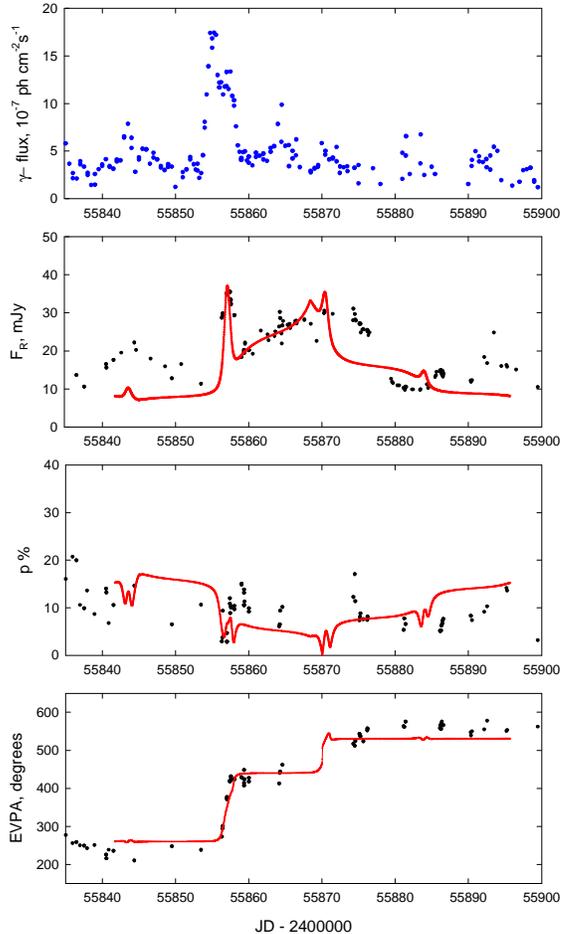}
        \caption{Enlarged part of Fig.~\ref{s5total} with model fitting.}
\label{model}
\end{center} 
\end{figure}

Figure \ref{model} shows the time evolution of the observational parameters ($\gamma$-ray and optical flux, degree of polarization and positional angle) within the model discussed above.  Values of the adjustable parameters are listed in Table~\ref{tab_1}. Several of the trends of the data resemble similar features in the model flux and polarization curves. Particularly, we see a succession of quasi-periodic flares before and after the main outburst (marked with vertical arrows in Fig~\ref{s5total2011}) and fast oscillation of the degree of polarization around the main flare. This conforms with the model presented here.  Turbulence may play a role in modulating the behavior (e.g., the discrepant flux and degree of polarization near MJD~55675) beyond the capabilities of the simplified model proposed here.

\begin{deluxetable*}{cccccccccccccccc}
\singlespace
\tablecolumns{16}
\tablecaption{\bf Fitting parameters for the photometric and polarimetric behavior of S5~0716+71 in 2011 October-November.\label{tab_1}}
\tabletypesize{\footnotesize}
\tablehead{
\colhead{$\theta\degr$}&\colhead{$\zeta\degr$}&\colhead{$p_\mathrm{jet}$,\%}&\colhead{$k$}&\colhead{$\Gamma$}&\colhead{$r$}&\colhead{$P_\mathrm{obs}$}&\colhead{$P_\mathrm{src}$}&\colhead{$\tau$}&\colhead{$T_0$}&\colhead{$R1$}&\colhead{$R2$}&\colhead{$a$}&\colhead{$\eta$}&\colhead{$t_\mathrm{del}$}&\colhead{$\Delta \chi\degr$} \\
\colhead{(1)}&\colhead{(2)}&\colhead{(3)}&\colhead{(4)}&\colhead{(5)}&\colhead{(6)}&\colhead{(7)}&\colhead{(8)}&\colhead{(9)}&\colhead{(10)}&\colhead{(11)}&\colhead{(12)}&\colhead{(13)}&\colhead{(14)}&\colhead{(15)}&\colhead{(16)}
}
\startdata
5.8&7.5&22&1.8&10&0.0095&13.2&1.5&0.47&0.11&0.06&132&1.1&1.50&0.47&10 \\
\enddata
\tablecomments{Units of $r$ are parsecs, $P_\mathrm{obs}$ -- days, $P_\mathrm{src}$ -- years. $\tau$, $T_0$ and $t_\mathrm{del}$ are fractions of $P_\mathrm{src}$.}
\end{deluxetable*}

The $\gamma$-ray outburst started $1\fd$4 before the optical, which corresponds to $\sim 0.05$~pc distance in the source frame. The emergence of knot K3 at the same time as the optical/$\gamma$-ray outburst leads to the conclusion that these events were co-spatial. However, we note that our model uses a lower Lorentz factor of the shock wave ($\Gamma=10$) and a wider viewing angle ($\theta=5.8\degr$) than that obtained from the radio images ($\Gamma=21$). This could reflect acceleration of the plasma flow in the jet downstream from the regions where the optical/$\gamma$-ray outburst took place \citep[see, e.g.,][]{2009ApJ...706.1253H}. 

\begin{acknowledgements}
We acknowledge the referee's useful comments and suggestions. This work was partly supported by Russian RFBR grants 12-02-00452, 12-02-31193 and by NASA Fermi Guest Investigator grants NNX-08AV65G, -11AO37G and -11AQ03G. The VLBA is operated by the National Radio Astronomy Observatory. The
National Radio Astronomy Observatory is a facility of the National Science
Foundation operated under cooperative agreement by Associated
Universities, Inc. The PRISM camera at Lowell Observatory
was developed by K. Janes et al. at BU and Lowell Observatory,
with funding from the NSF, BU, and Lowell Observatory.
The Liverpool Telescope is operated on
the island of La Palma by Liverpool John Moores University
in the Spanish Observatorio del Roque de los Muchachos of
the Instituto de Astrofisica de Canarias, with financial support
from the UK Science and Technology Facilities Council.
\end{acknowledgements}

\end{document}